\documentstyle[11pt]{article}

\renewcommand{\vec}[1]{{\bf #1}}       
\def\beq{\begin{eqnarray}}    
\def\eeq{\end{eqnarray}}      

\def\pa{\partial}                       
\def\al{\alpha}
\def\be{\beta}

\def\ga{\gamma}

\def\ep{\epsilon}

\def\na{\nabla}

\def\si{\sigma}

\begin{document}

\begin{center}

{\large\bf
On the interaction of massive spinor particles with external}
\vskip 2mm
{\large\bf electromagnetic and torsion fields}


\vskip 5mm
\setcounter{page}1
\renewcommand{\thefootnote}{\arabic{footnote}}
\setcounter{footnote}0

{\bf Lewis H. Ryder} $^{a}$
\footnote{ e-mail address: l.h.ryder@uk.ac.uk}
$\,\,$
and
$\,\,$
{\bf Ilya L. Shapiro} $^{b,\,c}$
\footnote{ e-mail address: shapiro@ibitipoca.fisica.ufjf.br}

\vskip 3mm

{\sl $^{a}\,\,$
Physics Laboratory, University of Kent,	Canterbury, Kent, CT2 7NR, UK}

\vskip 1mm

{\sl $^{b}\,\,$
Departamento de Fisica, Universidade Federal de Juiz de Fora,
36036-330, MG -- Brazil }

\vskip 1mm

{\sl $^{c}\,\,$
Tomsk State Pedagogical University, 634041, Tomsk, Russia }

\end{center}

\vskip 4mm

\noindent
{\large \it Abstract.} {\small \it
We explore the Dirac equation
in external electromagnetic and torsion fields.
Motivated by the previous study of quantum field theory
in an external torsion field we include a nonminimal interaction
of the spinor field with torsion.  As a consequence
the torsion axial vector and the electromagnetic potential
enter the action in a similar form.
The existence of an extra local symmetry is emphasized
and the Foldy-Wouthuysen transformation is performed to an accuracy of
next to the leading order. We also discuss the motion of a
classical test particle in a constant torsion field.}

\vskip 8mm

\noindent
{\bf Introduction.}
$\,\,\,$
Research in the field of gravity with torsion and especially
the interaction of torsion with a spinor field
has attracted attention for a long time \cite{dat,hehl,aud,rum,rump}
(see \cite{hehl-review} for a recent review of gravity with torsion).
Recently interest in theories with torsion has
increased because of the success of the formal development
of string theory \cite{GSW} which is (together with its modifications and
generalizations) nowadays regarded as the main candidate for the
unique description of all quantum fields.
String theory predicts the existence of the space-time torsion.
Probably this was the reason why in
recent years there has been an increasing interest in possible
physical effects of torsion
\cite{babush,hammond,doma3,lamme,sinryd}.
Most of these works discuss the effects of classical or quantum matter
fields on an external torsion background.

The study of the renormalization of quantum field theory in an external
gravitational field with torsion \cite{bush,book} has shown that
the interaction of matter fields with torsion has special features.
If the theory includes Yukawa interaction, then the
renormalizability requires nonminimal interaction of torsion
with both spinors and scalars. Thus we arrive at the necessity
to introduce some new nonminimal parameters which characterizes
such an interaction.
The considerations of \cite{babush,hammond,lamme,sinryd}
were based on the nonrelativistic limit of the Dirac equation --
that is on the generalization of the Pauli equation for the case  of
an external electromagnetic and torsion fields.
The main purpose of the present paper is to derive the next to the
leading order corrections to the Pauli equation in the framework of the
Foldy-Wouthuysen transformation.
We also
demonstrate some global symmetry which holds for the Dirac spinor
in external electromagnetic and torsion fields.

The paper is organized in the following way. In the next section we
introduce basic notations for the gravity with
torsion. Some additional symmetry which holds for the spinor field
coupled to torsion is established.
In section 3 the details of the Foldy--Wouthuysen transformation
are presented. The equations of motion for spinning particles are
discussed in section 4 and section 5 contains our conclusions.

\vskip 3mm
\noindent
{\bf 2. Quantum fields in an external gravitational field with torsion.}
$\,\,\,$
Let us start with the basic notions of gravity with torsion.
All our notations correspond to those in \cite{book}.
The metric $g_{\mu\nu}$ and torsion $T^\alpha_{\;\beta\gamma}$
are independent characteristics of  space - time.
When torsion is present, the covariant derivative $\tilde{\nabla}$
is based on the nonsymmetric connection
$\tilde{\Gamma}^\alpha_{\;\beta\gamma}$
\beq
\tilde{\Gamma}^\alpha_{\;\beta\gamma} -
\tilde{\Gamma}^\alpha_{\;\gamma\beta} =
T^\alpha_{\;\beta\gamma}
\label{tor}
\eeq
The metricity condition
$\tilde{\nabla}_\mu g_{\alpha\beta} = 0$ enables one to express
 the connection through the metric and torsion in a unique way 
\cite{book}.

It is convenient to divide the torsion field into three irreducible
components:
i) the vector $T_{\beta} = T^\alpha_{\,\;\beta\alpha}\,,\,\,$
ii) the axial vector
$\;S^{\nu} = \epsilon^{\alpha\beta\mu\nu}T_{\alpha\beta\mu}\;\,$ and
iii) the tensor
$\;q^\alpha_{\;\beta\gamma}\;$, which satisfies conditions
$q^\alpha_{\;\beta\alpha} = 0\,\,;\,\,
\epsilon^{\alpha\beta\mu\nu}q_{\alpha\beta\mu} =0$.
The torsion field can be always expressed through these new fields as
\beq
T_{\alpha\beta\mu} = \frac{1}{3} \left( T_{\beta}\,g_{\alpha\mu} -
T_{\mu}\,g_{\alpha\beta} \right) -
\frac{1}{6}\, 
\varepsilon_{\alpha\beta\mu\nu}\, S^{\nu} + q_{\alpha\beta\mu}
\label{irr}
\eeq
We remark that the string effective action contains the completely
antisymmetric torsion which can be described by the axial vector $S_\al$.

To construct the actions of the matter fields in an external
gravitational field with torsion one imposes the principles of
locality, and general covariance and require the symmetries of the given
theory
(like gauge invariance for the QED or SM) in flat space-time to hold
for the generalized theory in curved space-time with torsion, and
forbid the introduction of the new parameters with the
dimension of inverse mass.
It turns out that these demands fix the form of the
action apart from the values of some new parameters which
remain arbitrary. These new quantities are called nonminimal parameters
and parameters of the vacuum action.

Along with the nonminimal scheme described above,
there is a (more traditional) minimal one. According to it
the partial derivatives $\pa_\mu$
are substituted by the
covariant ones $\tilde{\nabla}_\mu$, the flat
metric $\eta^{\mu\nu}$ by $g^{\mu\nu}$ and the
volume element $d^4x$ by the covariant expression $d^4x\sqrt{-g}$.
For the Dirac spinor this procedure leads to the expression (see, for
example, \cite{hehl,book})
\beq
S_{\frac12 , min} = i\,\int d^4x\sqrt{-g}\,{\bar \psi}\, \left(\,
\ga^\al \,{\pa}_\al + {i}\eta_1\,\ga_5\ga^\al\,S_\al -
i\eta_2 \ga^\al\,T_\al
- im\, \right)\,\psi
\label{dirac}
\eeq
with $\eta_1=- 1/8$ and $\eta_2=0$. Therefore minimally
the Dirac spinor interacts only with the $S_\al$ part of the
torsion tensor (\ref{irr}).

The nonminimal interaction is essentially more complicated.
In the spinor sector the
nonminimal interaction is given by (\ref{dirac}) with an arbitrary
$\eta_{1,2}$. The introduction of the
nonminimal parameters in the matter field sector of GUT and
also the vacuum action is dictated by the requirements of
renormalizability, because corresponding counterterms
appear even if the starting theory had minimal interactions only
\cite{bush}. One has to remember, however, that among all the
nonminimal parameters only the ones which have relation to the
axial piece $S_\mu$ of the torsion tensor are essential \cite{bush,book}.
Since we are only interested in the torsion effects,
in the rest of this paper we will keep the metric flat.

In this letter we will be concerned by
the action of the nonminimally coupled Dirac spinor
in external electromagnetic and torsion fields (from this point
we consider a flat metric only).
We notice, that since the trace of torsion $T_\mu$ and the
electromagnetic
field $A_\mu$ appear only in the combination $eA_\mu + \eta T_\mu$,
it is natural to denote this combination simply $eA_\mu$ and therefore we
can drop $T_\mu$ everywhere. Moreover we change $\eta_1$ to $\eta$.
\beq
S_{1/2}= i\,\int d^4x \,{\bar \psi}\, \left[
\ga^\al \,\left( {\pa}_\al - ieA_\al + i\eta\ga_5\,S_\al\,\right)
- im \right]\,\psi
\label{diraconly}
\eeq

It is well known that for the spinor field in an external
electromagnetic field the gauge transformation leaves the action
invariant. In the case of a massless action
(\ref{diraconly}) this transformation can be
generalized in the following way:
$$
\psi \rightarrow \psi' = \psi\,e^{\al + \be \ga_5}
,\,\,\,\,\,\,\,\,\,\,\,\,\,\,\,\,\,\,\,\,\,\,\,\,\,\,
{\bar {\psi}} \rightarrow {\bar {\psi}}' = {\bar {\psi}}
\,e^{- \al + \be \ga_5}\,,
$$
\beq
A_\mu \rightarrow A_\mu ' = A_\mu - \frac{1}{e}\, \pa_\mu\al
\,,\,\,\,\,\,\,\,\,\,\,\,\,\,\,\,\,\,\,\,\,\,\,\,\,\,\,
S_\mu \rightarrow S_\mu ' = S_\mu - \frac{1}{\eta}\, \pa_\mu\be
\label{trans}
\eeq
where $\,\al = \al(x)\,$ and $\,\be = \be(x)\,$ are scalar and
pseudoscalar parameters of transformation.
Despite this symmetry is softly broken by the mass of the spinor field
in (\ref{diraconly}), it turns out to be very useful. In particular,
if the torsion field is considered as an external background,
this extra symmetry explains the special "gauge invariant" form
of divergences of the effective action of torsion which was observed
in \cite{buodsh,book} for the case of massless spinor field.

If we intend to consider
the propagating torsion, such a symmetry puts rigid restrictions
on the form of the action for the $S_\mu$ field \cite{belsh}. The
torsion field is supposed to interact with both massive and massless
spinors. Then one can show that
the only possible form of the torsion action which is compatible
with the softly broken symmetry (\ref{trans}) and leads to
simultaneously leads to the renormalizable quantum theory is
\beq
S_{ts} = \int d^4x\,\left[-\frac14\,S_{\mu\nu}^2 + \frac12\,M_{ts}^2
\,S_{\mu}^2 \right]\, + \,\mbox{(interaction terms)}
\label{toract}
\eeq
where $\,S_{\mu\nu} = \pa_\mu S_{\nu}  - \pa_\nu S_{\mu}\,$
and $M_{ts}$ is the mass of the torsion field.

Thus the only one propagating part of the torsion axial vector is
the transversal one, while the appearance of the longitudinal part
of $\,S_{\mu}\,$
is due to the massive terms in (\ref{toract}) and
(\ref{diraconly}), and the equation of motion for this
part of $\,S_{\mu}\,$ is algebraic with the corresponding part of spin
tensor as a source. Indeed this doesn't put any restrictions
on the value of this part which can be nonzero and even
large in a very early Universe. On the other hand,
even if the longitudinal component of torsion could be large
at the moments close to the Big Bang; during the inflationary period
it could decrease in many orders of magnitude \cite{buodsh} and
moreover the inhomogeneities of such a primeval torsion could
become irrelevant even at the cosmic distances. Therefore,
despite the propagation of torsion is related to the transversal
part of $\,S_{\mu}\,$, there is nothing inconsistent to suppose
that the primeval torsion has a nonzero longitudinal component.
In particular, it can be constant field of an arbitrary (including
the spacelike) configuration.

This supposition is in fact important for the discussion of the
results of recent paper \cite{doma3}.
The main hypotheses of \cite{doma3} is that there is a torsion
background radiation $\,S_\mu = (S_0,{\vec S})\,$ with a very small but
nonzero constant $\,{\vec S}$. Then integrating over the
spinor fields in the theory with an action (\ref{diraconly})
the Lagrangian of external fields $\,A_\mu ,\,S_\mu$ results in the
form $\,\ep^{\mu\nu\al\be}\,A_\mu S_\nu F_{\al\be}\,$. This Lagrangian
is exactly the one which is necessary to explain the observed anisotropy
of the cosmological electromagnetic propagation \cite{nodrol}.
As one can easily see, the above Lagrangian is gauge invariant
(with respect to
the usual transformation of the electromagnetic potential $A_{\mu}$)
if only $\,S_\mu\,$ is a longitudinal axial vector.
Therefore the existence of this component is a necessary constituent
of the mentioned explanation.

Another form of the torsion field arises
if we suppose that the torsion tensor doesn't violate the
isotropy of the Universe. Then in our cosmological frame the time
component of the vector $S_{\mu}$ should dominate. The existing
inflationary cosmological model with torsion predicts that during
the period of inflation the torsion field decreased exponentially
\cite{buodsh}. On the other hand during this period all possible
inhomogeneities of torsion have to disappear since we observe some
small part of the Universe. Therefore today we would observe only a very
weak and weakly dependent in space and time components of the
axial vector $S_\mu$.

The question whether the torsion field really exists
can be solved only experimentally, so it is important to study the
features of different torsion configurations.
Thus it is worth of studying the possible manifestations
of torsion at low energies, where there is a possibility of making very
precise measurements.

\vskip 3mm
\noindent
{\bf 3. Foldy-Wouthuysen transformation.}
$\,\,\,$
In this section we shall derive the Foldy-Wouthuysen transformation
for the Dirac spinor in external torsion and electromagnetic fields.
The initial Hamiltonian corresponding to (\ref{diraconly}),
has the form:
\beq
H = \be m + {\cal E} + {\cal G}
\label{inham}
\eeq
where
$\,\,
 {\cal E} = e\,A_0 - \eta\,\ga_5\,{\vec {\al}}\,{\vec S}
\,,\,\,\,\,\,{\cal G} = {\vec {\al}}
\, \left({\vec p} - e {\vec A} \right)
+ \eta\,\ga_5\,S_0\,\,\,\,$
are the even and odd parts of the expression,
$S_\mu = \left(S_0, {\vec S}\right)$ is the axial piece of the 
torsion tensor (\ref{irr})and we have used the
standard representation of the gamma-matrices and the units
$c={\hbar} = 1 $.

Our purpose is to find a unitary transformation which separates
"small" and "large" components of the Dirac spinor.
We use a general prescription
\beq
H' = e^{i{\cal S}}
\,\left( H - i \,\partial_t\right)\,e^{-i{\cal S}}
\label{gener}
\eeq
where ${\cal S}$ has to be chosen in an appropriate way. We shall find
${\cal S}$ and $H'$ in a form of the nonrelativistic expansion,
and thus will take ${\cal S}$ to be of order $1/m$.
Then we can use standard result (see, for example, \cite{BD})
$$
H' = H + i \,\left[{\cal S},H \right] -
\frac12\,\left[{\cal S}, \left[{\cal S},H \right] \right] -
\frac{i}{6}\,\left[{\cal S}, \left[{\cal S}, \left[{\cal S},H
\right]  \right] \right] +
\frac{1}{24}\,\left[{\cal S}, \left[{\cal S}, \left[{\cal S},
\left[{\cal S},H \right] \right]  \right] \right] -
$$
\beq
- {\dot {\cal S}}
- \frac{i}{2}\,\left[{\cal S}, {\dot {\cal S}} \right] +
\frac{1}{6}\,\left[{\cal S}, \left[{\cal S}, {\dot {\cal S}} \right]\right]
+ ...
\label{expan}
\eeq
One can easily see that ${\cal E}$ and ${\cal G}$ given above
(anti)commute with $\be$ in a standard way
$\,\,{\cal E}\,\be = \be\, {\cal E}
\,,\,\,{\cal G} \,\be = - \be\, {\cal G}\,$
and therefore one can safely use the standard prescription for
the lowest-order approximation for ${\cal S}$:
$\,{\cal S} = - \frac{i}{2m}\,\be \, {\cal G}\,$.
\beq
H' = \be m + {\cal E}' + {\cal G}'
\label{1ham}
\eeq
where ${\cal G}'$ is of order $1/m$, and one has to perform
second Foldy-Wouthuysen transformation with
${\cal S}' = - \frac{i}{2m}\,\be {\cal G}'$. This leads to the
\beq
H'' = \be m + {\cal E}' + {\cal G}''
\label{2ham}
\eeq
with ${\cal G}'' \approx 1/m^2$; and then a third
Foldy-Wouthuysen transformation with
${\cal S}'' = - \frac{i}{2m}\,\be {\cal G}''$ removes odd operators
in the given order of the nonrelativistic
expansion, so that we finally obtain the usual result
\beq
H''' = \be\,\left( m + \frac{1}{2m}\,{\cal G}^2 -
\frac{1}{8m^3}\,{\cal G}^4 \right) + {\cal E}
- \frac{1}{8m^2}\,\left[{\cal G}, \left(
\left[{\cal G}, {\cal E}\right] + i\,{\dot {\cal G}} \right)\right]
\label{usual}
\eeq
Substituting here ${\cal E}$ and ${\cal G}$
after some algebra we arrive at the final form of the Hamiltonian
$$
H''' = \be \left[ m + \frac{1}{2m}
\left( {\vec p}
- e {\vec A} + \eta S_0 {\vec {\si}} \right)^2 -
\frac{1}{8m^3}\,{\vec p}^4 \right] + eA_0
- \eta \left( {\vec {\si}}\cdot{\vec S} \right)
- \frac{e}{2m} {\vec {\si}} \cdot {\vec B} - \frac{\eta^2}{m}\,\be S_0^2 -
$$$$
- \frac{e}{8m^2}\,\left[ div {\vec E} +
i{\vec {\si}}\cdot {\rm curl} {\vec E}
+ 2 {\vec {\si}}\cdot\,\left({\vec E} \times {\vec p}\right) \right]
+\frac{\eta}{8m^2}\,\left[
- {\vec {\si}}\cdot\nabla {\dot S}_0 + \{p_i,\{p^i,
\left({\vec {\si}}\cdot{\vec S}\right)\}\} -
\right.
$$
\beq
\left.
- 2 \left\{ \left({\vec T}\cdot{\vec p}\right),
\left({\vec {\si}}\cdot{\vec p}\right) \right]
- 2\, {\rm curl}{\vec S}\cdot {\vec p} +
2i \left({\vec {\si}}\cdot\nabla\right)
\left({\vec S}\cdot  {\vec p}\right)
+ 2i\,\left({\na}{\vec S}\right)\,\left({\vec {\si}}\cdot{\vec p}\right)
\right\}
\label{separado}
\eeq
where we use standard notation $\left\{A,B \right\} = AB + BA$.
One can indeed proceed in this way and get separated Hamiltonian
with any given accuracy in $\,1/m$.
The first five terms of (\ref{separado})
reproduce the Pauli-like equation
with torsion derived in \cite{babush} (see also \cite{hammond}).
The next term is the next-to-the-leading order
nonrelativistic correction for the case of external electromagnetic
field without torsion. The remaining terms are
torsion-dependent corrections to the Pauli-like equation \cite{babush}
(see also \cite{hammond,lamme}).
In those terms we follow the same system of approximation as is
standard for the electromagnetic case \cite{BD}; that is we keep
terms $\,\,(kinetic \,\, energy)^3\,$ and
$\,(kinetic \,\, energy)^2\cdot(potential \,\, energy)$.

Further simplifications are possible if we are
interested in relic cosmological torsion. In this
case we have to keep only
the constant components of the axial $S_\mu$.
Then the effects of torsion will be:
i) a small correction to the potential energy of the spinor field,
(which sometimes looks just like a correction to the mass) and
ii) the appearance of a new gauge-invariant interaction
spin-momentum term in the Hamiltonian.

\vskip 3mm
\noindent
{\bf 4. Motion of spinning particle on the constant torsion background.}
$\,\,\,$
The above expression (\ref{separado})
is rather cumbersome and difficult to work with. However, one can
make some general observations.
In particular, at this order of approximation the corrections
to the Pauli-like equations which are related to the constant
part $S_0$ of the axial piece of torsion enter the Hamiltonian only via
the expression
${\vec {\pi}} = {\vec p} - e {\vec A} + \eta S_0 {\vec \si}$. Therefore,
according to the result of \cite{babush} the second order correction
does not modify the equation of motion for the classical spinning
particle.

Consider now the motion of a spinning particle in the space with
torsion but without any electromagnetic field.
The physical degrees of freedom of the particle are its coordinate
${\vec {x}}$ and its spin ${\vec {\si}}$.
For the reasons explained in the Introduction we are mainly
interested in the cases of constant axial
$S_\mu = (S_0, {\vec {S}})$. In this case the corresponding equations of
\cite{babush} have the form:
\beq
\frac{d{\vec {v}}}{dt}
= - \eta \,{\vec {S}}\,({\vec {v}}\cdot {\vec {\si}})
- \frac{\eta S_0}{c}\,\frac{d{\vec {\si}}}{dt}
\label{eq1}
\eeq
\beq
\frac{d{\vec {\si}}}{dt} =
+ \frac{2\eta}{\hbar}\,\left[{\vec {S}}\times {\vec {\si}}\right]
- \frac{2\eta S_0}{\hbar c}\,\left[{\vec {v}}\times {\vec {\si}}\right]
\label{eq2}
\eeq
Consider first the case suggested in \cite{doma3}, when
$S_0 = 0$ so that only ${\vec {S}}$ is present. Since ${\vec S} = const$,
we can safely put $S_{1,2}=0$. The solution for spin
can be easily found to be
\beq
{\si}_3 = {\vec {\si}}_{30} = const,\,\,\,\,\,\,\,\,\,\,\,\,
\si_1 = \rho\,\cos \left( \frac{2\eta S_3t}{\hbar}\right)
,\,\,\,\,\,\,\,\,\,\,\,\,
\si_2 = \rho\,\sin \left( \frac{2\eta S_3t}{\hbar}\right)
\label{sol-spin}
\eeq
where $\rho = \sqrt{\si_{10}^2 + \si_{20}^2}$.
For the first two components of the velocity we have
$\,\,v_1=v_{10}=const,\,\,\,v_2=v_{20}=const,\,\,$
 but the solution for $v_3$ turns out to be complicated. For $\si_3 = 0$
one finds
\beq
v_3(t) =  v_{30} - \frac{\hbar \rho v_{20}}{2m}
\,-\, \frac{\rho \hbar v_{10}}{2m}
\,\sin \left( \frac{2\eta S_3t}{\hbar}\right)\, +
\,\frac{\rho \hbar v_{20}}{2m}
\,\cos \left( \frac{2\eta S_3t}{\hbar} \right)
\label{sol-skor1}
\eeq
and for $\si_3 \neq 0$ the solution is
$$
v_3(t) = \left[\, v_{30}\, + \,
\frac{\rho\hbar\,\left(\si_3 v_{10}\hbar -
2mv_{20}\right)}{4m^2 + \hbar^2\si_3^2}\,\right]
\;e^{-\frac{\eta S_3 \si_3}{m}\,t}\, -
$$
\beq
-\,\frac{\rho\hbar}{4m^2 + \hbar^2\si_3^2}\;\left[
\, \left(\si_3 v_{10}\hbar - 2mv_{20}\right)
\,\cos \left( \frac{2\eta S_3t}{\hbar}\right)
\,+\, \left(\si_3 v_{20}\hbar + 2mv_{10}\right)
\,\sin \left( \frac{2\eta S_3t}{\hbar}\right) \,\right]
\label{sol-skor2}
\eeq
Physically such a solution means i) precession of the spin around the
direction of ${\vec S}$ and ii) oscillation of the particle velocity
in this same direction
is accompanied (for $\si_3\neq 0$ )
by the exponential damping of the initial velocity in
this direction. We remark that the value of the
relic torsion field should be very weak so that a very precise
experiments will be necessary to measure this (probably extremely
slow) precession, oscillation and damping.

Consider finally the
other special case ${\vec {S}} = 0$, which is the form
of the torsion field motivated by isotropic cosmological models.
Then the equations of motion have a
form different from the ones occurring in a magnetic field
\beq
\frac{d{\vec {v}}}{dt}
= - \frac{\eta S_0}{c}\,\frac{d{\vec {\si}}}{dt} =
\frac{2\eta^2 S_0^2}{c \hbar}\,
\left[{\vec {v}}\times {\vec {\si}}\right]
\label{eq3}
\eeq
Despite the fact that those
 equations formally look like a nonlinear system
of equations for six unknowns, they can be integrated immediately
if we notice that the time variations of the variables do not affect
the vector product. Hence the general solutions are
$$
{\vec {v}}(t) = {\vec {v}}_0 +
\left[{\vec {v}}_0\times {\vec {\si}}_0\right]\,
\frac{2\eta^2S_0^2}{c\hbar}\,t
$$
\beq
{\vec {\si}}(t) = {\vec {\si}}_0 -
\left[{\vec {v}}_0\times {\vec {\si}}_0\right]\,
\frac{2\eta
 S_0}{\hbar}\,t
\label{solu}
\eeq
The first equation indicates that the motion of such a particle is
a motion with constant acceleration. This is possible, in the presence
of torsion, for electrically neutral particles with spin.
A remark is in order here.
The torsion field is supposed to act on the spin of particles
but not on their angular momentum \cite{hehl}. Therefore a motion
like (\ref{solu}) will occur for individual electrons or other
particles with spin as well as for macroscopic bodies with
a nonzero overall spin orientation but it does not occur for
the (charged or neutral) bodies with a random orientation of spins.
We note that the possibility of the accelerating motion of
the spinning particles in an external torsion field has been discusses
in \cite{sinryd}.

\vskip 3mm
\noindent
{\bf 5. Conclusions.}
$\,\,\,$
We have considered some aspects of the motion of Dirac particles
nonminimally coupled to external torsion and electromagnetic fields.
The usual
electrodynamic gauge invariance becomes generalized to (\ref{trans}).
This symmetry transformation has
one extra pseudoscalar parameter compared with standard gauge invariance
but in this extended form it holds only in the massless sector
of the action.

Despite the fact that the Dirac equation with an additional torsion
background
is more complicated than it is in the purely electromagnetic case,
it admits the Foldy-Wouthuysen transformation which separates the "large"
and "small" components of the Dirac spinor field. The higher order
corrections to the Pauli equation do not involve the constant timelike
component $S_0$ of the torsion dual axial.

The motion of classical spinning particles in an external torsion
field shows some original features if the torsion field has
only a time component $S_0$. In this case the motion with constant
acceleration is possible for an electrically neutral particle.

\vskip 3mm
\noindent
{\bf Acknowledgments.}
$\,\,\,$
One of the authors (I.L.Sh.) is grateful to Departamento de F\'{i}sica,
Universidade Federal de Juiz de Fora for warm hospitality.
His work was supported in part by the CNPq (Brazil) and by
Russian Foundation for Basic Research under project No.96-02-16017.

\end{document}